\pdfoutput=1
\RequirePackage{ifpdf}
\ifpdf % We~are running pdfTeX in pdf mode
\documentclass[pdftex]{sigma}
\else
\documentclass{sigma}
\fi

\numberwithin{equation}{section}

\usepackage{mathrsfs}
\usepackage{stmaryrd}

\def\Z#1{\mathbb{Z}_2^{#1}}
\def\g{\mathfrak{g}}
\def\s{\mathfrak{s}}
\def\pX{\pmb{X}}
\def\pY{\pmb{Y}}
\def\pZ{\pmb{Z}}
\def\pW{\pmb{W}}

\begin{document}
%\allowdisplaybreaks

\newcommand{\arXivNumber}{2103.10638}

\renewcommand{\PaperNumber}{071}

\FirstPageHeading

\ShortArticleName{$\mathbb{Z}_2^3$-Graded Extensions of Lie Superalgebras and Superconformal Quantum Mechanics}

\ArticleName{$\boldsymbol{\mathbb{Z}_2^3}$-Graded Extensions of Lie Superalgebras\\ and Superconformal Quantum Mechanics}

\Author{Shunya DOI and Naruhiko AIZAWA}
\AuthorNameForHeading{S.~Doi and N.~Aizawa}
\Address{Department of Physical Science, Osaka Prefecture University, \\
Nakamozu Campus, Sakai, Osaka 599-8531, Japan}
\Email{\href{mailto:s_s.doi@p.s.osakafu-u.ac.jp}{s\_s.doi@p.s.osakafu-u.ac.jp}, \href{mailto:aizawa@p.s.osakafu-u.ac.jp}{aizawa@p.s.osakafu-u.ac.jp}}

\ArticleDates{Received March 24, 2021, in final form July 14, 2021; Published online July 20, 2021}

\Abstract{Quantum mechanical systems whose symmetry is given by $\mathbb{Z}_2^3$-graded version of superconformal algebra are introduced. This is done by finding a realization of a $\mathbb{Z}_2^3$-graded Lie superalgebra in terms of a standard Lie superalgebra and the Clifford algebra. The realization allows us to map many models of superconformal quantum mechanics (SCQM) to their $\mathbb{Z}_2^3$-graded extensions. It is observed that for the simplest SCQM with $\mathfrak{osp}(1|2)$ symmetry there exist two inequivalent $\mathbb{Z}_2^3$-graded extensions. Applying the standard prescription of conformal quantum mechanics, spectrum of the SCQMs with the $\mathbb{Z}_2^3$-graded $\mathfrak{osp}(1|2)$ symmetry is analyzed. It is shown that many models of SCQM can be extended to $\mathbb{Z}_2^n$-graded setting.}


\Classification{17B75; 17B81; 81R12}

\section{Introduction}

In the recent works \cite{AAd2,AAD, BruDup}, $\Z{n}$-graded extensions of supersymmetric quantum mechanics (SQM) were introduced and their properties were investigated ($\Z{n}$ denotes the direct product of $n$ copies of the Abelian group $\mathbb{Z}_2$).
They are a quantum mechanical realization of $\Z{n}$-graded version of supersymmetry algebra introduced by Bruce \cite{Bruce} (see also \cite{LR}), i.e.,
the Hamiltonian is a matrix differential operator acting on a $\Z{n}$-graded Hilbert space and the symmetry is given by a~$\Z{n}$-graded Lie superalgebra.
As a $\Z{n}$-graded Lie superalgebra (see Appendix for definition) is an extension of Lie superalgebra to more complex grading structure \cite{Ree,rw1,rw2,sch1},
the $\Z{n}$-graded SQM is a natural generalization of the standard SQM.
It is observed in \cite{AAd2, AAD} that the $\Z{n}$-graded SQM is constructed by a combination of the standard SQM and Clifford algebras.
In fact, it is known that a tensor product of a Clifford algebra and a standard Lie superalgebra realizes a $\Z{n}$-graded Lie superalgebra \cite{NAcl, rw2}.
Such realization is not unique since for a given Lie superalgebra there exist some distinct ways of tensoring Clifford algebras.
Usually, the distinct tensoring produces inequivalent $\Z{n}$-graded extensions of the Lie superalgebra. However, it can happen that those $\Z{n}$-graded extensions are identical and the different tensoring produces inequivalent representations of a single $\Z{n}$-graded Lie superalgebra. This is the case of $\Z{n}$-graded SQM studied in \cite{AAd2} where tensor product of a standard SQM and a sequence of Clifford algebras gives inequivalent representations of the $\Z{n}$-graded version of supersymmetry algebra.

The realizations in \cite{AAd2} are restricted to the standard SQM and it is not clear that it can be applicable to other Lie superalgebras.
On the other hand, the realization presented in \cite{AAD} is applied to a larger class of Lie superalgebras though it produce only $\Z{2}$-graded extensions.
Thus one can use it to define a $\Z{2}$-graded extension of superconformal quantum mechanics (SCQM).
It is shown that by this realization many models of the standard SCQM are mapped to their $\Z{2}$-graded extension.
The simplest case, $\Z{2}$-graded $\mathfrak{osp}(1|2)$ SCQM, is investigated in some detail~\cite{AAD} and an abstract representation theory of the $\Z{2}$-graded $\mathfrak{osp}(1|2)$ is developed in \cite{KANA} where the richness of irreducible representations of the $\Z{2}$-graded $\mathfrak{osp}(1|2)$ is observed.

As a continuation of the works on quantum mechanical realizations of $\Z{n}$-graded Lie superalgebras, in the present work we explore $\Z{n}$-graded version of SCQM and present models of $\Z{3}$-graded SCQM explicitly.
Although we focus on $\Z{3}$-graded SCQM,
models of $ \Z{n}$-graded SCQM for any $n$ are also introduced.
In fact, our result is more general since we start with a new way of mapping a standard Lie superalgebra to its higher graded version.
This means that higher graded extensions of physically relevant algebras such as super-Poincar\'e, super-Schr\"odinger etc are also obtained in our formalism.

The present work is motivated by the recent renewed interest in $\Z{n}$-graded superalgebras in physics and mathematics.
In physics side, they give a new symmetry different from the ones generated by Lie algebras and superalgebras.
Here we mention only some of them.
It was found that symmetries of some differential equations such as L\'evy-Leblond equation (non-relativistic Dirac equation) are generated by $\Z{2}$-graded Lie superalgebras \cite{AKTT1,AKTT2}.
Some supersymmetric classical theories are extended to $\Z{2}$-graded setting \cite{AKTcl,AKTqu,brusigma}.
It is shown that non-trivial physics can be detected in the multiparticle sector of the $\Z{2}$-graded SQM \cite{Topp}.
$\Z{2}$-Graded version of spacetime symmetries are proposed by several authors, e.g.,~\cite{tol2}.
In mathematics side, $\Z{n}$-graded supergeometry which is an extension of supergeometry on supermanifolds, is studied extensively, see, e.g.,~\cite{Pon}.
More exhaustive list of references of physical and mathematical aspects of $\Z{n}$-graded Lie superalgebras is found in~\cite{KANA}.

We organise this paper as follows:
We start Section~\ref{SEC:pre} with the definition of $\Z{3}$-graded Lie superalgebra.
Then we review briefly the results of~\cite{AAd2} on $\Z{n}$-graded SQM.
An emphasis is put on the fact that there exists a sequence of inequivalent models of $\Z{n}$-graded SQM for a given standard SQM because we also consider the sequence of $\Z{3}$-graded SCQM in this work.
In Section~\ref{SEC:CL4} the first member of the sequence (there are three members), ${\rm Cl}(4)$ model, is presented.
We give a realization of a $\Z{3}$-graded Lie superalgebra in terms of the Clifford algebra ${\rm Cl}(4)$ and a standard Lie superalgebra.
This realization is applied to $\mathfrak{osp}(1|2)$ SCQM, then we obtain its $\Z{3}$-graded extension.
The spectrum of $\Z{3}$-graded $\mathfrak{osp}(1|2)$ SCQM is investigated by employing the standard procedure of conformal quantum mechanics.
In Section~\ref{SEC:Cl6} other two members of the sequence, for which ${\rm Cl}(6)$ is used, are considered and it will be shown that one of them is irrelevant as it does not give an irreducible representation of $\Z{3}$-graded $\mathfrak{osp}(1|2)$.
For the relevant one which defines an another $\Z{3}$-graded extension of $\mathfrak{osp}(1|2)$,
the same analysis as ${\rm Cl}(4)$ case is repeated.
We close the paper with some remarks in Section~\ref{SEC:CR}.

\section{Preliminaries} \label{SEC:pre}

\subsection[Z23-graded Lie superalgebras]{$\boldsymbol{\Z{3}}$-graded Lie superalgebras}

We define the $\Z{3}$-graded Lie superalgebra according to \cite{rw1,rw2}.
Let $ \vec{a} = (a_1, a_2, a_3)$, $\vec{b} = (b_1, b_2, b_3) $ be elements of $\Z{3}$.
Here we regard an element of $\Z{3}$ as a three-dimensional vector and their
sum and inner product are computed in modulus 2
\begin{equation*}
 \vec{a}+ \vec{b} = (a_1+b_1, a_2+b_2, a_3+b_3), \qquad
 \vec{a}\cdot \vec{b} = \sum_{k=1}^3 a_k b_k.
\end{equation*}
We also introduce the parity of $ \vec{a} $ defined by
\begin{equation*}
 |\vec{a}| := \sum_{k=1}^3 a_k \mod 2.
\end{equation*}
Consider a complex vector space $\g$ consisting of eight subspaces each of which is labelled by an element of $\Z{3}$:
\begin{equation*}
 \g = \g_{(0,0,0)} \oplus \g_{(0,0,1)} \oplus \g_{(0,1,0)} \oplus \g_{(1,0,0)} \oplus
 \g_{(0,1,1)} \oplus \g_{(1,0,1)} \oplus \g_{(1,1,0)} \oplus \g_{(1,1,1)}.
\end{equation*}
The vector space $\g$ is refereed to as a $\Z{3}$-graded Lie superalgebra if its elements are closed in commutator or anticommutator which is chosen according to the following rule
\begin{equation}
 \llbracket X_{\vec{a}}, X_{\vec{b}} \rrbracket
 :=
 \begin{cases}
 [X_{\vec{a}}, X_{\vec{b}} ], & \vec{a}\cdot\vec{b} = 0,
 \vspace{1mm} \\
 \{X_{\vec{a}}, X_{\vec{b}} \}, & \vec{a}\cdot\vec{b} =1,
 \end{cases}
 \qquad
 X_{\vec{a}} \in \g_{\vec{a}}, \qquad X_{\vec{b}} \in \g_{\vec{b}}.
 \label{doublebracket}
\end{equation}
We use $\llbracket \cdot, \cdot \rrbracket $ as a unified notation of commutator and anticommutator.
See appendix for more rigorous definition of $\Z{n}$-graded Lie superalgebras.

\subsection[A sequence of Z2n-graded SQM]{A sequence of $\boldsymbol{\Z{n}}$-graded SQM}

We review the results of \cite{AAd2} briefly as the present work is an algebraic generalization of them.
The $\Z{n}$-graded SQM is defined as a realization of $\Z{n}$-graded supersymmetry algebra in a $\Z{n}$-graded Hilbert space.
The $\Z{n}$-graded supersymmetry algebra consists of one Hamiltonian, $2^{n-1}$~supercharges of parity 1 and $2^{n-2}\big(2^{n-1}-1\big)$ central elements of parity~0.

It was shown that a tensor product of ${\mathcal N}=1$ standard SQM and a complex irreducible representation (irrep) of the Clifford algebra ${\rm Cl}(2m)$ can give the realization which define a~model of the $\Z{n}$-graded SQM.
The ${\mathcal N}=1$ standard SQM is generated by one supercharge~$Q$ and its defining relations are given by
\begin{equation*}
 \{Q, Q\} = 2H, \qquad [H, Q] = 0.
\end{equation*}
Both $Q$ and $H$ are $2\times 2$ matrix differential operators acting on $\mathbb{Z}_2$-graded Hilbert space.

The Clifford algebra ${\rm Cl}(2m)$ is generated by $\gamma_j$ $(j=1,2,\dots, 2m)$ subject to the conditions
\begin{equation*}
 \{\gamma_j, \gamma_k \} = 2\delta_{jk}.
\end{equation*}
The Hermitian complex irrep of ${\rm Cl}(2m)$ is $2^m$-dimensional and
given explicitly as follows \cite{CaRoTo, Okubo}
\begin{align}
 \gamma_1 &= \sigma_1^{\otimes m},
 \qquad
 \gamma_j = \sigma_1^{\otimes (m-j+1)} \otimes \sigma_3 \otimes \mathbb{I}_2^{\otimes (j-2)}, \qquad 2 \leq j \leq m,
 \nonumber \\
 \tilde{\gamma}_j &:= \gamma_{j+m} = \sigma_1^{\otimes(m-j)} \otimes \sigma_2 \otimes \mathbb{I}_2^{\otimes(j-1)},
 \qquad 1 \leq j \leq m, \label{Clrep}
\end{align}
where $\sigma_k$ is the Pauli matrix and $\mathbb{I}_2$ denotes the $2\times 2$ identity matrix.
Therefore, a model of $\Z{n}$-graded SQM is a set of $2^{m+1}$-dimensional matrix differential operators.

For a fixed value of $n$, one may have a sequence of inequivalent models of $\Z{n}$-graded SQM by tensoring the standard SQM and the following sequence of the Clifford algebra
\begin{equation}
 {\rm Cl}(2(n-1)), \quad {\rm Cl}(2n), \quad {\rm Cl}(2(n+1)),\quad \dots, \quad {\rm Cl}(2^n-2).
 \label{ZnSeq}
\end{equation}
For instance, we have five distinct models of $\Z{4}$-graded SQM from the Clifford algebras
\begin{equation*}
 {\rm Cl}(6), \quad {\rm Cl}(8), \quad {\rm Cl}(10), \quad {\rm Cl}(12), \quad {\rm Cl}(14).
\end{equation*}
The difference in the models is the number of linearly independent central elements.
The $\Z{n}$-graded supersymmetry algebra has a lot of central elements.
Some of the central elements are realized as dependent operators unless the Clifford algebra of the maximal dimension in the above sequence is used.
Lower the dimension of the Clifford algebra, more central elements are realized as dependent operators.

In the next two sections, we show the existence of a sequence of realizations of $\Z{3}$-graded Lie superalgebra and by which one may introduce models of $\Z{3}$-graded SCQM.

\section[Cl(4) model of Z23-graded SCQM]{$\boldsymbol{{\rm Cl}(4)}$ model of $\boldsymbol{\Z{3}}$-graded SCQM} \label{SEC:CL4}

In this and the following sections, we deal with $\Z{3}$-graded Lie superalgebras and $\Z{3}$-graded SCQM.
Setting $n=3$ in~\eqref{ZnSeq}, we see that the sequence has only two Clifford algebras: $ {\rm Cl}(2(n-1)) = {\rm Cl}(4)$, ${\rm Cl}(2n)={\rm Cl}(2^n-2) = {\rm Cl}(6)$.
However, realizations for ${\rm Cl}(2n)$ and $ {\rm Cl}(2^n-2)$ considered in \cite{AAd2} are not identical.
We thus explore three cases, one for ${\rm Cl}(4)$ and two for ${\rm Cl}(6)$.
It will then turn out that, contrary to $\Z{n}$-graded SQM, we have two inequivalent $\Z{3}$-graded extensions of $\mathfrak{osp}(1|2)$.
In this section, we focus on ${\rm Cl}(4)$.

In the sequel, we denote a standard ($\mathbb{Z}_2$-graded) Lie superalgebra by $\s $ and its even and odd subspaces by $\s_0 $ and $ \s_1$, respectively.
We use a Hermitian representation of $\s$ to realize a $\Z{3}$-graded Lie superalgebra.
Recalling that $ |\vec{a}| = 0 $ or $1$ for $\vec{a} \in \Z{3}$,
we denote a Hermitian matrix representing an element of $\s_{|\vec{a}|}$ by $ X_{|\vec{a}|}$ and suppose its size is $ 2m \times 2m$.

\subsection[Cl(4) realization of Z23-graded Lie superalgebra]{$\boldsymbol{{\rm Cl}(4)}$ realization of $\boldsymbol{\Z{3}}$-graded Lie superalgebra}\label{SEC:Cl4Alg}

The irrep \eqref{Clrep} for ${\rm Cl}(4)$ is given by
\begin{gather*}
 	\gamma_{1} = \sigma_{1} \otimes \sigma_{1} ,\qquad
 	\gamma_{2} = \sigma_{1} \otimes \sigma_{3},\qquad
 	\gamma_{3} =\sigma_{1} \otimes \sigma_{2},\qquad
 	\gamma_{4} = \sigma_{2} \otimes \mathbb{I}_{2}.
\end{gather*}
Let $ \Gamma $ be a matrix subject to
\begin{equation}
 [X_0, \Gamma] = 0, \qquad
 \{X_{1},\Gamma\}=0,\qquad \Gamma^{2}=\mathbb{I}_{2m}, \qquad
 \forall\, X_{|\vec{a}|} \in \s_{|\vec{a}|}. \label{GammaCond}
\end{equation}
Then the matrices defined by
\begin{gather} \label{Cl4mapping}
 \pX_{\vec{a}} = {\rm i}^{f(\vec{a})} \gamma_1^{a_1} \gamma_2^{a_2}
 \otimes X_{|\vec{a}|} \Gamma^{a_1+a_2},
\qquad
 f(\vec{a}) := a_1 a_2 + |\vec{a}| (a_1 + a_2) \mod 2
\end{gather}
are Hermitian and define a $\Z{3}$-graded Lie superalgebra.
More explicitly, $ \pX_{\vec{a}}$ is given by
\begin{alignat*}{4}
&\pX_{(0,0,0)}=\mathbb{I}_{4} \otimes X_{0}, \qquad &&
\pX_{(1,0,0)}={\rm i}\gamma_{1} \otimes X_{1}\Gamma ,\qquad&&
\pX_{(0,1,0)}={\rm i}\gamma_{2} \otimes X_{1}\Gamma ,& \\
&\pX_{(0,0,1)}=\mathbb{I}_{4} \otimes X_{1} , \qquad&&
\pX_{(1,1,1)} ={\rm i}\gamma_{1}\gamma_{2} \otimes X_{1} ,\qquad &&
\pX_{(1,1,0)}={\rm i}\gamma_{1}\gamma_{2} \otimes X_{0},& \\
&\pX_{(1,0,1)}=\gamma_{1} \otimes X_{0}\Gamma , \qquad &&
\pX_{(0,1,1)}=\gamma_{2} \otimes X_{0}\Gamma. \qquad &&&
\end{alignat*}
It is immediate to see $ \pX_{\vec{a}}$ is Hermitian
\begin{gather*}
 (\pX_{\vec{a}})^{\dagger} =
 (-{\rm i})^{f(\vec{a})} \gamma_2^{a_2} \gamma_1^{a_1} \otimes
 \Gamma^{a_1+a_2} X_{|\vec{a}|}
 = (-1)^{f(\vec{a})+a_1 a_2 + |\vec{a}| (a_1+a_2)} \pX_{\vec{a}} = \pX_{\vec{a}}.
\end{gather*}
To verify the $\Z{3}$-graded Lie superalgebra structure,
we need to prove the closure in (anti)com\-mu\-ta\-tor and graded Jacobi relations~\eqref{GJdef}.
This will be done by showing that the $\Z{3}$-graded commutators and Jacobi relations are reduced to those for the Lie superalgebra $\s$.
It is not difficult to see the (anti)commutators (see~\eqref{GLB}) are computed as
\begin{align}
 \llbracket \pX_{\vec{a}}, \pX_{\vec{b}} \rrbracket
 &= X_{\vec{a}} X_{\vec{b}} - (-1)^{\vec{a}\cdot \vec{b}} X_{\vec{b}}X_{\vec{a}}
 \nonumber \\
 &= (-1)^{a_2b_1+(a_1+a_2)|\vec{b}|} {\rm i}^{f(\vec{a})+f(\vec{b})}
 \gamma_1^{a_1+b_1} \gamma_2^{a_2+b_2} \otimes
 \langle X_{|\vec{a}|},X_{|\vec{b}|} \rangle \Gamma^{a_1+a_2+b_1+b_2}, \label{com22to2}
\end{align}
where
\[
 \langle X_{|\vec{a}|},X_{|\vec{b}|} \rangle := X_{|\vec{a}|}X_{|\vec{b}|}-(-1)^{|\vec{a}||\vec{b}|}X_{|\vec{b}|} X_{|\vec{a}|}
\]
is the (anti)commutator of the Lie superalgebra $\s$.
Writing the (anti)commutation relations of~$\s$ in the form
\begin{equation*}
 \langle X_{|\vec{a}|},X_{|\vec{b}|} \rangle = {\rm i}^{1-|\vec{a}| |\vec{b}|} X_{|\vec{a}|+|\vec{b}|} = {\rm i}^{1-|\vec{a}| |\vec{b}|} X_{|\vec{a}+\vec{b}|},
\end{equation*}
\eqref{com22to2} yields
\begin{equation*}
 \llbracket \pX_{\vec{a}}, \pX_{\vec{b}} \rrbracket
 =
 (-1)^{a_2b_1+(a_1+a_2)|\vec{b}|} {\rm i}^{1+f(\vec{a})+f(\vec{b})-f(\vec{a}+\vec{b})- |\vec{a}| |\vec{b}|} \pX_{\vec{a}+\vec{b}}
\end{equation*}
with
\begin{equation*}
 \pX_{\vec{a}+\vec{b}} = {\rm i}^{f(\vec{a}+\vec{b})} \gamma_1^{a_1+b_1} \gamma_2^{a_2+b_2}
 \otimes X_{|\vec{a}+\vec{b}|} \Gamma^{a_1+a_2+b_1+b_2}.
\end{equation*}
Therefore, $ \llbracket \pX_{\vec{a}}, \pX_{\vec{b}} \rrbracket \in \g_{\vec{a}+\vec{b}}$, i.e., closure of $\Z{3}$-graded (anti)commutator has been proved.

By the similar computation one may see
\begin{align*}
 (-1)^{\vec{a}\cdot\vec{c}} \llbracket \pX_{\vec{a}}, \llbracket \pX_{\vec{b}}, \pX_{\vec{c}} \rrbracket \rrbracket
=& (-1)^{\kappa} {\rm i}^{f(\vec{a})+f(\vec{b})+f(\vec{c})} \gamma_1^{a_1+b_1+c_1} \gamma_2^{a_2+b_2+c_2}
 \\
 & {}\otimes (-1)^{|\vec{a}| |\vec{c}|}
 \langle X_{|\vec{a}|}, \langle X_{|\vec{b}|}, X_{|\vec{c}|} \rangle \rangle
 \Gamma^{\sum_{k=1}^2(a_k + b_k + c_k)},
\end{align*}
where
\begin{gather*}
 \kappa := \sum_{k=1}^2(a_k b_k + b_k c_k + c_k a_k)
 + a_1 b_2 + b_1 c_2 + c_1 a_2
 \\
\hphantom{\kappa :=}{}
 + a_1 b_3 + b_1 c_3 + c_1 a_3 + a_2 b_3 + b_2 c_3 + c_2 a_3.
\end{gather*}
$\kappa$ is invariant under the cyclic permutation of $ a$, $b$, $c$.
This shows that the graded Jacobi relations are reduced to those for~$\s$.
It follows that the graded Jacobi identity holds true and the $\Z{3}$-graded Lie superalgebra structure has been proved.

The realization \eqref{Cl4mapping} is able to generalize to a realization of $\Z{n}$-graded Lie superalgebras by~$\mathfrak{s}$ and ${\rm Cl}(2(n-1))$:
\begin{gather}
	 \pX_{\vec{a}} = {\rm i}^{f(\vec{a})} \prod_{j=1}^{n-1}\gamma_{j}^{a_{j}} \otimes X_{|\vec{a}|}\Gamma^{\sum_{k=1}^{n-1}a_{k}},
	 \label{Cl4general}
	 \\
	 f(\vec{a}) = \sum_{k=1}^{n-2} a_{k}\prod_{l=k+1}^{n-1}a_{l} + |\vec{a}|\sum_{l=1}^{n-1}a_{l} \mod 2.\nonumber
\end{gather}
One may prove this in the same way as $\Z{3}$-graded Lie superalgebras so we do not present the proof.

\subsection[Cl(4) model of Z23-graded osp(1|2) SCQM]{$\boldsymbol{{\rm Cl}(4)}$ model of $\boldsymbol{\Z{3}}$-graded $\boldsymbol{\mathfrak{osp}(1|2)}$ SCQM} \label{SubSEC:CL4}

As shown in Section~\ref{SEC:Cl4Alg}, any Lie superalgebra satisfying the condition~\eqref{GammaCond} can be promoted to a~$\Z{3}$-graded superalgebra.
If one starts with a matrix differential operator realization of a~superconformal algebra, i.e., a model of SCQM, then one may obtain its $\Z{3}$-graded version.
Many models of SCQM have been obtained so far (see, e.g., \cite{FedIvaLec,Oka,Pa}).
Some of the models, e.g., the ones in \cite{ACKT, AKT}, satisfy the condition \eqref{GammaCond} so that we may have the $\Z{3}$-graded SCQM of ${\mathcal N}=2, 4, 8$ and so on.

Here we analyse the simplest example of $\Z{3}$-graded SCQM obtained from the $\mathfrak{osp}(1|2)$ superconformal algebra.
Let us consider the following realization of $\mathfrak{osp}(1|2)$ which is a ${\mathcal N}=1$ SCQM:
\begin{alignat}{4}
& Q= \frac{1}{\sqrt{2}} \left( \sigma_1 p - \sigma_2 \frac{\beta}{x} \right),
 \qquad&&
 S= \frac{x}{\sqrt{2}} \sigma_1,\qquad &&&
 \nonumber \\
& H= \frac{1}{2} \left( p^2 + \frac{\beta^2}{x^2} \right) \mathbb{I}_2 + \frac{\beta}{2x^2}\sigma_3,
\qquad && D = -\frac{1}{4} \{ x, p \} \mathbb{I}_2, \qquad && K = \frac{x^2}{2} \mathbb{I}_2, & \label{ospSCM}
\end{alignat}
where $ \beta $ is a coupling constant.
The non-vanishing relations of $\mathfrak{osp}(1|2)$ read as follows
\begin{alignat*}{4}
& [D, K]={\rm i}K, \qquad && [H, K] = 2{\rm i}D, \qquad && [D, H]= -{\rm i}H,& \nonumber \\
& \{Q, Q\}= 2H, \qquad && \{S, S\}= 2K, \qquad && \{Q, S\}=-2D,& \nonumber \\
& [D, Q]=-\frac{{\rm i}}{2} Q, \qquad && [D, S]= \frac{{\rm i}}{2} S,\qquad && [Q, K]= -{\rm i}S,& \nonumber \\
& [S, H]= {\rm i}Q. && && &
\end{alignat*}

One may immediately see that $\Gamma = \sigma_3 $ satisfies the condition \eqref{GammaCond}.
Thus by \eqref{Cl4mapping} we obtain twenty operators:
the diagonal degree $(0,0,0)$ operators are given by
\begin{equation}
 \pmb{H}_{000}=\mathbb{I}_{4} \otimes H,\qquad
 \pmb{D}_{000}=\mathbb{I}_{4} \otimes D,\qquad
 \pmb{K}_{000}=\mathbb{I}_{4} \otimes K. \label{Cl4000}
\end{equation}
Here and in the following sections we use a simplified notation $ \pX_{a_1a_2a_3} := \pX_{(a_1,a_2,a_3)}. $
The operator $ \pmb{H}_{000} $ is the Hamiltonian of the model and these three operators form the one-dimensional conformal algebra $ \mathfrak{so}(1,2) $.
The other parity even operators, which are not diagonal, are given by
\begin{gather*}
 \pmb{X}_{110}=i\gamma_{1}\gamma_{2} \otimes X,\qquad
 \pmb{X}_{101}=\gamma_{1} \otimes X\sigma_{3} ,\qquad
 \pmb{X}_{011}=\gamma_{2} \otimes X\sigma_{3},\qquad X= H, \, D, \, K,
\end{gather*}
and the parity odd ones are given by
\begin{alignat}{3}
& \pmb{Q}_{100}={\rm i}\gamma_{1} \otimes Q\sigma_{3}, \qquad&&
 \pmb{S}_{100} ={\rm i}\gamma_{1} \otimes S\sigma_{3}, & \nonumber \\
& \pmb{Q}_{010}={\rm i}\gamma_{2} \otimes Q\sigma_{3}, \qquad&&
 \pmb{S}_{010}={\rm i}\gamma_{2} \otimes S\sigma_{3},& \nonumber \\
 &\pmb{Q}_{001}=\mathbb{I}_{4} \otimes Q, \qquad&&
 \pmb{S}_{001}=\mathbb{I}_{4} \otimes S,& \nonumber \\
& \pmb{Q}_{111}={\rm i}\gamma_{1}\gamma_{2} \otimes Q, \qquad&&
 \pmb{S}_{111}={\rm i}\gamma_{1} \gamma_{2} \otimes S.&
 \label{Z3ospSCM1}
\end{alignat}
(Anti)commutator of these operators are closed and define a $\Z{3}$-graded extension of $\mathfrak{osp}(1|2)$ which we denote simply by $ {\mathcal G}_1$. We note that $ \dim {\mathcal G}_1=20 $.

For the range of $\beta$ where the potential is repulsive, the Hamiltonian $H$ in \eqref{ospSCM} has continuous spectrum.
It is known that the eigenfunctions of $H$ with the positive eigenvalue are plane wave normalizable, however, the zero energy state is not even plane wave normalizable~\cite{DFF}.
This property is inherited to the Hamiltonian $\pmb{H}_{000}$ of the $\Z{3}$-graded SCQM \eqref{Z3ospSCM1}.
In order to analyse the syetem \eqref{Z3ospSCM1} we follow the standard prescription of conformal mechanics. That is, the eigenspace of $\pmb{H}_{000} $ is not taken as the Hilbert space of the theory. Instead, the eigenspace of an operator which is a linear combination of $\pmb{H}_{000}$ and $\pmb{K}_{000}$ is chosen as the Hilbert space. We thus introduce the following operators
\begin{alignat*}{3}
& \pmb{R}_{\vec{a}}= \pmb{H}_{\vec{a}} + \pmb{K}_{\vec{a}},
\qquad &&
 \pmb{L}_{\vec{a}}^{\pm}= \frac{1}{2}( \pmb{K}_{\vec{a}}- \pmb{H}_{\vec{a}}) \pm {\rm i} \pmb{D}_{\vec{a}},& \\
& \pmb{a}_{\vec{a}}= \pmb{S}_{\vec{a}} + {\rm i} \pmb{Q}_{\vec{a}},
\qquad &&
 \pmb{a}_{\vec{a}}^{\dagger}= \pmb{S}_{\vec{a}} -{\rm i} \pmb{Q}_{\vec{a}}.&
\end{alignat*}
The diagonal operator $\pmb{R}_{000}$ is the new Hamiltonian and it has discrete eigenvalues due to the oscillator potential (see \eqref{ospSCM} and~\eqref{Cl4000}).
The eigenspace of $\pmb{R}_{000}$ is $ \mathscr{H} = L^2(\mathbb{R})\otimes \mathbb{C}^8$ which is taken to be the Hilbert space of the model.
The space $\mathscr{H}$ has a vector space decomposition according to the $\Z{3}$-degree
\begin{equation*}
 \mathscr{H} = \bigoplus_{\vec{a}\in \Z{3}} \mathscr{H}_{\vec{a}}.
\end{equation*}
The operators $ \pmb{a}_{\vec{a}}$, $\pmb{a}_{\vec{a}}^{\dagger}$, $\pmb{L}_{\vec{a}}^{\pm}$ generate the spectrum of $\pmb{R}_{000}$
\begin{gather}
 [\pmb{R}_{000}, \pmb{a}_{\vec{a}} ] = -\pmb{a}_{\vec{a}},
 \qquad
 \big[\pmb{R}_{000}, \pmb{a}_{\vec{a}}^{\dagger} \big] = -\pmb{a}_{\vec{a}}^{\dagger},
 \qquad
 \big[\pmb{R}_{000}, \pmb{L}_{\vec{a}}^{\pm}\big] = \pm 2 \pmb{L}_{\vec{a}}^{\pm},
 \label{CreationSp} \\
 \big\{ \pmb{a}_{\vec{a}}, \pmb{a}_{\vec{a}}^{\dagger} \big\} = 2 \pmb{R}_{000},\nonumber
\end{gather}
and the operators $ \pmb{a}_{\vec{a}}$, $\pmb{a}_{\vec{a}}^{\dagger}$, together with $F$ defined below, form a Klein deformed oscillator algebra
\begin{alignat*}{3}
& \big[\pmb{a}_{\vec{a}}, \pmb{a}_{\vec{a}}^{\dagger}\big]= \mathbb{I}_8 -2 \beta F,
 \qquad&& F:= \mathbb{I}_4 \otimes \sigma_3, &
 \\
& F^2= \mathbb{I}_{8}, \qquad && \{F, \pmb{a}_{\vec{a}} \}= \big\{F, \pmb{a}_{\vec{a}}^{\dagger} \big\} = 0.&
\end{alignat*}
It follows that
\begin{equation*}
 \pmb{R}_{000}
 = \pmb{a}_{\vec{a}}^{\dagger} \pmb{a}_{\vec{a}}
 + \frac{1}{2} (\mathbb{I}_8 -2 \beta F).
\end{equation*}
Thus the ground state is obtained by
\begin{equation*}
 \pmb{a}_{\vec{a}} \Psi(x)=0,
\end{equation*}
where
\begin{equation*}
 \Psi(x)=(\psi_{000}(x),\psi_{001}(x),\psi_{110}(x),\psi_{111}(x),\psi_{011}(x),\psi_{010}(x),\psi_{101}(x),\psi_{100}(x))^{\rm T}
 \in \mathscr{H}.
\end{equation*}
For all $ \pmb{a}_{\vec{a}} $ this condition is reduced to the set of relations for the components of $\Psi(x)$
\begin{alignat}{3}
& \left( \partial_x + x - \frac{\beta}{x} \right) \psi_{\vec{a}}(x) = 0,
 \qquad&&
 \vec{a} = (0,0,1), (1,1,1), (0,1,0), (1,0,0), & \label{GScondition1} \\
 &\left( \partial_x + x + \frac{\beta}{x} \right) \psi_{\vec{a}}(x)= 0,
 \qquad&&
 \vec{a} = (0,0,0), (1,1,0), (0,1,1), (1,0,1). & \label{GScondition2}
\end{alignat}
The solution of these equations are given by $ \psi_{\vec{a}}(x) = x^{\pm \beta} {\rm e}^{-x^2/2}$ and the normalizability of the functions are studied in detail in~\cite{ACKT}.
For $\beta > 1$ (repulsive potential) only one of them is normalizable so that the ground state is either
\begin{equation*}
 x^{\beta} {\rm e}^{-x^2/2}(0,C_1,0,C_2,0,C_3,0,C_4)^{\rm T}
\end{equation*}
with the energy $ \frac{1}{2}(1+2\beta) $ or
\begin{equation*}
 x^{-\beta} {\rm e}^{-x^2/2}(C_1,0,C_2,0,C_3,0,C_4,0)^{\rm T}
\end{equation*}
with the energy $ \frac{1}{2}(1-2\beta) $ where $ C_i$ is a constant.
Thus the ground state is four-fold degenerate and belongs to either parity odd or even subspaces of~$\mathscr{H}$.

The excited states with various $\Z{3}$-degree are obtained by repeated application of $\pmb{a}_{\vec{a}}^{\dagger}$ on the ground state and one may see from \eqref{CreationSp} that the operator $\pmb{R}_{(0,0,0)}$ has equally spaced spectrum.
We remark that no need to consider the action of $\pmb{L}_{\vec{a}}^+ $ because of the relation $ \big\{ \pmb{a}_{\vec{a}}^{\dagger}, \pmb{a}_{\vec{a}}^{\dagger} \big\} = 4 \pmb{L}^+_{\vec{a}}$.
The excited state is also four-fold degenerate.
This is verified as follows.
Let $\phi_{\vec{a}} \in \mathscr{H}_{\vec{a}} $ and $\phi_{\vec{b}} \in \mathscr{H}_{\vec{b}} $ be eigenfunctions of $\pmb{R}_{000}$ with the same eigenvalue.
Then $ \pmb{a}_{\vec{c}} \phi_{\vec{a}} $ equals to $\pmb{a}_{\vec{d}} \phi_{\vec{b}}$ up to a constant multiple if $ \vec{a} + \vec{c} = \vec{b} + \vec{d}. $
For instance, it is not difficult to see the following two functions in $\mathscr{H}_{(1,0,1)}$ are identical up to a constant
\begin{alignat*}{3}
 & \pmb{a}_{100} \psi_{001}, \qquad&& \psi_{001}=x^{\beta} {\rm e}^{-x^2/2} (0,C_1,0,0,0,0,0,0),& \\
 & \pmb{a}_{010} \psi_{111}, \qquad && \psi_{111}= x^{\beta} {\rm e}^{-x^2/2} (0,0,0,C_2,0,0,0,0).&
\end{alignat*}

 This ${\rm Cl}(4)$ model of $\Z{3}$-graded SCQM gives irreps of ${\mathcal G}_1$.
Recalling that the order of $\Z{3}$ is $\big|\Z{3}\big| = 8$, the minimal dimension of non-trivial irreps of ${\mathcal G}_1$ in a $ \Z{3}$-graded vector space is also eight which is the dimension of the ${\rm Cl}(4)$ model.

As mentioned at the end of Section~\ref{SEC:Cl4Alg}, we have a realization of $\Z{n}$-graded Lie superalgebras by $\mathfrak{s}$ and $ {\rm Cl}(2(n-1))$ where the condition \eqref{GammaCond} is required.
The condition is satisfied for $ \mathfrak{osp}(1|2)$ SCQM \eqref{ospSCM} as we have seen.
It is also satisfied for other models of SCQM.
For instance, $ \Gamma = \sigma_3, \sigma_3 \otimes \mathbb{I}_2 $ and
$ \left(\begin{smallmatrix}
 \mathbb{I}_8 & 0 \\ 0 & \mathbb{I}_8
\end{smallmatrix}\right) $ for the $\mathfrak{osp}(2|2)$, $D(2,1;\alpha) $ and $ F(4)$ models, respectively (see \cite{ACKT, AKT} for the models).
These models are promoted to their $\Z{n}$-graded version by using~\eqref{Cl4general}.
Thus, there exists various ${\rm Cl}(2(n-1))$ models of $\Z{n}$-graded SCQM and we may analyze its properties in a manner similar to this section.

\section[Cl(6) models of Z23-graded SCQM]{$\boldsymbol{{\rm Cl}(6)}$ models of $\boldsymbol{\Z{3}}$-graded SCQM} \label{SEC:Cl6}

In this section, we explore the models of $\Z{3}$-graded SCQM obtained via ${\rm Cl}(6)$ in a way similar to the ones via ${\rm Cl}(4)$.
As already mentioned, we investigate two realizations of $\Z{3}$-graded Lie superalgebras via ${\rm Cl}(2n)$ and ${\rm Cl}(2^n-2)$. These two Clifford algebras are degenerate for $n=3$, however the way of realizing the $\Z{3}$-graded Lie superalgebras are not identical.

The irrep \eqref{Clrep} for ${\rm Cl}(6)$, which is common for ${\rm Cl}(2n)$ and ${\rm Cl}(2^n-2)$, is given by
\begin{alignat*}{4}
&\gamma_{1}= \sigma_{1} \otimes \sigma_{1} \otimes \sigma_{1}, \qquad&&
\gamma_{2}= \sigma_{1} \otimes \sigma_{1} \otimes \sigma_{3}, \qquad&&
\gamma_{3}= \sigma_{1} \otimes \sigma_{3} \otimes \mathbb{I}_{2}, &\nonumber \\
&\gamma_{4}= \sigma_{1} \otimes \sigma_{1} \otimes \sigma_{2}, \qquad &&
\gamma_{5}= \sigma_{1} \otimes \sigma_{2} \otimes \mathbb{I}_{2},\qquad &&
\gamma_{6}= \sigma_{2} \otimes \mathbb{I}_{2} \otimes \mathbb{I}_{2}.&
\end{alignat*}

\subsection[Cl(2n) model]{$\boldsymbol{{\rm Cl}(2n)}$ model} \label{SEC:Cl2n}

The realization of $\Z{n}$-graded Lie superalgebra in terms of ${\rm Cl}(2n)$ and a ordinary Lie superalgebra~$\s$ is given in~\cite{NAcl}.
Thus we are able to use the result to investigate a model of $\Z{3}$-graded SCQM.
For $n=3$, the realization of $\Z{3}$-graded Lie superalgebra by $ X_a \in \s_a$ reads as follows
\begin{alignat*}{5}
&\pmb{X}_{000}=\mathbb{I}_{8} \otimes X_{0}, && && && &\nonumber \\
&\pmb{X}_{100}=\gamma_{1} \otimes X_{1}, \qquad && \pmb{X}_{010}=\gamma_{2} \otimes X_{1}, \qquad&&
\pmb{X}_{001}=\gamma_{3} \otimes X_{1}, \qquad&&
\pmb{X}_{111}={\rm i}\gamma_{1}\gamma_{2}\gamma_{3} \otimes X_{1},&\nonumber \\
& \pmb{X}_{110}={\rm i}\gamma_{1}\gamma_{2} \otimes X_{0}, \qquad\!&&
\pmb{X}_{101}={\rm i}\gamma_{1}\gamma_{3} \otimes X_{0},\qquad\! &&
\pmb{X}_{011}={\rm i}\gamma_{2}\gamma_{3} \otimes X_{0}.&& &
\end{alignat*}
Contrast to ${\rm Cl}(4)$, there is no condition like \eqref{GammaCond} so that any models of SCQM can be extended to $\Z{3}$-grading by this realization.
We consider again the $ \mathfrak{osp}(1|2)$ model \eqref{ospSCM} as the simplest example.
The $\Z{3}$-graded SCQM so obtained is the set of matrix differential operators
\begin{alignat}{4}
& \pmb{X}_{000}=\mathbb{I}_{8} \otimes X,\qquad && && & \nonumber \\
&\pmb{Q}_{100}=\gamma_{1} \otimes Q, \qquad &&
\pmb{S}_{100} =\gamma_{1} \otimes S, \qquad &&& \nonumber \\
&\pmb{Q}_{010}=\gamma_{2} \otimes Q,\qquad &&
\pmb{S}_{010} =\gamma_{2} \otimes S, \qquad &&& \nonumber \\
&\pmb{Q}_{001}=\gamma_{3} \otimes Q, \qquad &&
\pmb{S}_{001} =\gamma_{3} \otimes S, \qquad &&& \nonumber \\
&\pmb{Q}_{111}={\rm i}\gamma_{1} \gamma_{2} \gamma_{3} \otimes Q, \qquad&&
\pmb{S}_{111}={\rm i}\gamma_{1} \gamma_{2} \gamma_{3} \otimes S, \qquad &&& \nonumber \\
&\pmb{X}_{110}={\rm i}\gamma_{1} \gamma_{2} \otimes X, \qquad &&
\pmb{X}_{101}={\rm i}\gamma_{1} \gamma_{3} \otimes X, \qquad&&
\pmb{X}_{011}={\rm i}\gamma_{2} \gamma_{3} \otimes X &\label{Cl2nSCQM}
\end{alignat}
with $X = H, D, K $.

It is not difficult to see that these twenty operators form an closed algebra whose (anti)com\-mu\-ta\-tion relations are identical to the ones for ${\rm Cl}(4)$ model \eqref{Z3ospSCM1}.
Namely, the operators in \eqref{Cl2nSCQM} give 16-dimensional representation of $ {\mathcal G}_1$.
However, this is a reducible representation of $ {\mathcal G}_1$.
To see this, let $F(\mathbb{R})$ be a space of complex valued functions on a real line and
$ \mathscr{H} = F(\mathbb{R})\otimes \mathbb{C}^{16}$.
The operators \eqref{Cl2nSCQM} act on $\mathscr{H}$ and it is readily seen from the explicit form of the operators that the following subspaces $\mathscr{H}_1$ and $\mathscr{H}_2$ are invariant under the action of \eqref{Cl2nSCQM}:
\begin{gather*}
 \mathscr{H} = \mathscr{H}_1 \oplus \mathscr{H}_2,\\
 \mathscr{H}_1 = (\psi_{000},0,\psi_{110},0,\psi_{011},0,\psi_{101},0,0,\psi_{001},0,\psi_{111},0,\psi_{010},0,\psi_{100})^{\rm T},\\
 \mathscr{H}_2 =(0,\psi_{000},0,\psi_{110},0,\psi_{011},0,\psi_{101},\psi_{001},0,\psi_{111},0,\psi_{010},0,\psi_{100},0)^{\rm T}.
\end{gather*}
This shows that the operators \eqref{Cl2nSCQM} are a reducible representation of $ {\mathcal G}_1$.

Although the combination of $ {\rm Cl}(2n) $ and $ \mathfrak{osp}(1|2)$ are not physically relevant for $n=3$. This does not implies other models such as $ D(2,1;\alpha), F(4)$ are irrelevant, either.
At least, one may see the existence of various models of $ \Z{n}$-graded SCQM.
Because the realization given in \cite{NAcl} does not require any further conditions like \eqref{GammaCond} so than any models of standard SCQM can be promoted to their $\Z{n}$-graded counterparts. Physical relevance of those models should be studied case by case.

As already mentioned, some inequivalent realizations of $\Z{n}$-graded Lie superalgebras in terms of the Clifford algebra and a standard Lie superalgebra are known.
However, the present example elucidates not all such realizations are suitable for physical applications.
We need to find an appropriate one to discuss physical problems.
Therefore, general study of realizations of $\Z{n}$-graded Lie superalgebra by ordinary superalgebras is an important research problem.

\subsection[Cl(2n-2) model]{$\boldsymbol{{\rm Cl}(2^n-2)}$ model}

A non-trivial ${\rm Cl}(6)$ model is obtained by an analogy of $\Z{n}$-extension of SQM considered in~\cite{AAd2}.
First, we introduce the following ordering into the parity odd elements of $\Z{3}$:
\begin{gather*}
\vec{a}_{0} = (1,1,1), \qquad
\vec{a}_{1} = (1,0,0), \qquad
\vec{a}_{2} = (0,1,0), \qquad
\vec{a}_{3} = (0,0,1).
\end{gather*}
Define the following Hermitian matrices
\begin{gather}
\pmb{X}_{(0,0,0)} = \mathbb{I}_8 \otimes X_{0}, \nonumber \\
\pmb{X}_{\mu} = \gamma_{\mu} \otimes X_{1} , \nonumber \\
\pmb{X}_{\mu \nu} = {\rm i}^{1-\vec{a}_{\mu}\cdot\vec{a}_{\nu}}\gamma_{\mu}\gamma_{\nu} \otimes X_{0} ,\qquad \mu<\nu, \nonumber \\
\pmb{X}_{\mu \nu \rho} = {\rm i} \gamma_{\mu}\gamma_{\nu}\gamma_{\rho} \otimes X_{1} ,\qquad \mu < \nu <\rho, \nonumber \\
\pmb{X}_{0123} = {\rm i}\gamma_{1}\gamma_{2}\gamma_{3} \otimes X_{0}, \label{Cl6real2}
\end{gather}
where the Greek indices run from 0 to 3 and $ \gamma_{0}=\mathbb{I}_{8}. $
The suffix $(0,0,0)$ of $ \pmb{X}_{(0,0,0)} $ denotes its $\Z{3}$-degree where the original notation is restored to avoid confusion.
The $ \Z{3}$-degree of the matrices with the Greek indices is determined as follows
\begin{alignat*}{3}
& \deg(\pmb{X}_{\mu})= \vec{a}_{\mu}, \qquad&&
 \deg(\pmb{X}_{\mu\nu})= \vec{a}_{\mu} + \vec{a}_{\nu}, &\nonumber \\
& \deg(\pmb{X}_{\mu\nu\rho})= \vec{a}_{\mu} + \vec{a}_{\nu} + \vec{a}_{\rho},
 \qquad &&
 \deg(\pmb{X}_{0123})= (0,0,0).&
\end{alignat*}
With this assignment of $\Z{3}$-degree the matrix operators of \eqref{Cl6real2} define a $\Z{3}$-graded Lie superalgebra.
This is verified by observing that the $\Z{3}$-graded (anti)commutators are reduced to those for a standard superalgebra.
In order to see this we write the relations of the superalgebra~$\mathfrak{s}$ as follows
\[
 [X_0, Y_0] = {\rm i} Z_0, \qquad
 [X_0, Y_1] = {\rm i} Z_1, \qquad
 \{X_1, Y_1 \} = W_0.
\]
By definition $ \llbracket \pX_{(0,0,0)}, \pY \rrbracket = [\pX_{(0,0,0)}, \pY] $ for any $\pY$. One may compute the commutator as follows{\allowdisplaybreaks
\begin{gather*}
 [ \pX_{(0,0,0)}, \pY_{(0,0,0)} ] = \mathbb{I}_8 \otimes [X_0, Y_0]
 = {\rm i} \pZ_{(0,0,0)},\\
 [ \pX_{(0,0,0)}, \pY_{\mu} ] = \gamma_{\mu} \otimes [X_0, Y_1] = {\rm i} \pZ_{\mu}, \\
 [ \pX_{(0,0,0)}, \pY_{\mu\nu} ] = {\rm i}^{1-\vec{a}_{\mu}\cdot\vec{a}_{\nu}} \gamma_{\mu} \gamma_{\nu} \otimes [X_0, Y_0] = {\rm i} \pZ_{\mu\nu},\\
 [ \pX_{(0,0,0)}, \pY_{\mu\nu\rho} ] = {\rm i} \gamma_{\mu} \gamma_{\nu} \gamma_{\rho} \otimes [X_0, Y_1] = {\rm i} \pZ_{\mu\nu\rho},\\
 [ \pX_{(0,0,0)}, \pY_{0123} ] = {\rm i} \gamma_1 \gamma_2 \gamma_3 \otimes [X_0, Y_0] = {\rm i} \pZ_{0123}.
\end{gather*}
Similarly for $ \pX_{0123}$
\begin{gather*}
 [\pX_{0123}, \pY_{\mu}] = {\rm i}\gamma_1 \gamma_2 \gamma_3 \gamma_{\mu} \otimes [X_0, Y_1]
 = \sum_{\nu,\rho,\sigma} f_{\mu \nu\rho\sigma} \pZ_{\nu\rho\sigma},
 \\
 [\pX_{0123}, \pY_{\mu\nu}] = -{\rm i}^{-\vec{a}_{\mu}\cdot\vec{a}_{\nu}} \gamma_1 \gamma_2 \gamma_3 \gamma_{\mu} \gamma_{\nu} \otimes [X_0, Y_0]
 = {\rm i}\sum_{\rho, \sigma} g_{\mu\nu\rho\sigma} \pZ_{\rho\sigma},
\\
 [\pX_{0123}, \pY_{\mu\nu\rho} ] = -\gamma_1 \gamma_2 \gamma_3 \gamma_{\mu} \gamma_{\nu} \gamma_{\rho} \otimes [X_0, Y_1]
 = {\rm i} \sum_{\sigma} h_{\mu\nu\rho\sigma} \pZ_{\sigma},\\
 [\pX_{0123}, \pY_{0123}] = \mathbb{I}_8 \otimes [X_0, Y_0] = {\rm i} \pZ_{(0,0,0)},
\end{gather*}
where the structure constants are given as follows
\begin{alignat*}{3}
& f_{0123} = f_{1023} = f_{3012} = 1, \qquad && f_{2013}=-1,& \nonumber \\
& g_{0123}= g_{0312} = g_{1203} = g_{2301} = 1, \qquad && g_{0213}= g_{1302} = -1,&
 \\
& h_{0123}= h_{0231} = h_{1230} = 1, \qquad && h_{0132}= -1,&
\end{alignat*}
and the others are zero.}

Other (anti)commutation relations are more involved
(\eqref{doublebracket}, see also \eqref{com22to2}):
\begin{gather}
 \llbracket \pX_{\mu}, \pY_{\nu} \rrbracket = \pX_{\mu} \pY_{\nu} -(-1)^{\vec{a}_{\mu}\cdot \vec{a}_{\nu}} \pY_{\nu} \pX_{\mu}
 = \gamma_{\mu}\gamma_{\nu} \otimes \{ X_{1},Y_{1} \}
 = {\rm i}^{-1+\vec{a}_{\mu} \cdot \vec{a}_{\nu}} \pW_{\mu\nu},\nonumber \\
 \llbracket \pX_{\mu},\pY_{\nu \rho} \rrbracket = \pX_{\mu} \pY_{\nu \rho} -(-1)^{\vec{a}_{\mu} \cdot (\vec{a}_{\nu}+\vec{a}_{\rho})} \pY_{\nu \rho} \pX_{\mu}
 = {\rm i}^{1-\vec{a}_{\nu}\cdot\vec{a}_{\rho}}\gamma_{\mu}\gamma_{\nu}\gamma_{\rho} \otimes [X_{1},Y_{0}]
 = {\rm i}^{1-\vec{a}_{\nu}\cdot\vec{a}_{\rho}} \pW_{\mu\nu\rho},
 \nonumber \\
 \llbracket \pX_{\mu}, \pY_{\nu \rho \sigma} \rrbracket
 = \pX_{\mu} \pY_{\nu \rho \sigma} -(-1)^{\vec{a}_{\mu}\cdot (\vec{a}_{\nu}+\vec{a}_{\rho}+\vec{a}_{\sigma})} \pY_{\nu \rho \sigma} \pX_{\mu}\nonumber\\
 \hphantom{\llbracket \pX_{\mu}, \pY_{\nu \rho \sigma} \rrbracket}{}
 = {\rm i} \gamma_{\mu}\gamma_{\nu}\gamma_{\rho}\gamma_{\sigma} \otimes \{X_{1},Y_{1} \}
 = \pW_{\mu\nu\rho\sigma}, \label{ClosureCl6no1}
 \end{gather}
where we introduced $ W_1 \in \mathfrak{s}_1 $ by $[X_1, Y_0] = {\rm i} W_1$ and $ \pW_{\mu\nu\rho} = {\rm i}\gamma_{\mu}\gamma_{\nu} \gamma_{\rho} \otimes W_1$.
The indices of $\pW$ on the right-hand side of~\eqref{ClosureCl6no1} do not always respect the restriction given in \eqref{Cl6real2}.
Such $\pW$ is converted to the one in \eqref{Cl6real2} by the following relations:
\begin{alignat*}{3}
& \pW_{\mu\mu}= \pW_{(0,0,0)},\qquad &&
 \pW_{\mu\nu} = (-1)^{1-\vec{a}_{\mu} \cdot \vec{a}_{\nu}} \pW_{\nu\mu},
 \quad \mu \neq \nu,& \\
& \pW_{\mu\mu\rho}= {\rm i} \pW_{\rho}, \qquad && \pW_{\mu\nu\mu}= -{\rm i} \pW_{\nu},& \nonumber \\
& \pW_{\mu\mu\nu\rho}= \pW_{\mu\nu\rho\mu} = (-{\rm i})^{\vec{a}_{\nu} \cdot \vec{a}_{\rho}} \pW_{\nu\rho}, \qquad &&
 \pW_{\mu\nu\mu\rho}= -(-{\rm i})^{\vec{a}_{\nu} \cdot \vec{a}_{\rho}} \pW_{\nu\rho},& \\
& \pW_{\nu\mu\mu\rho}= {\rm i}^{\vec{a}_{\nu} \cdot \vec{a}_{\rho}} \pW_{\nu\rho},
 \qquad &&
 \pW_{\nu\mu\rho\mu}= -{\rm i}^{\vec{a}_{\nu} \cdot \vec{a}_{\rho}} \pW_{\nu\rho},&
\end{alignat*}
and if all the indices are different value, then
\begin{gather}
 \pW_{\mu\nu\rho} = (-1)^{1-\vec{a}_{\mu}\cdot\vec{a}_{\nu}} \pW_{\nu\mu\rho}
 = (-1)^{\vec{a}_{\mu}\cdot( \vec{a}_{\nu} + \vec{a}_{\rho})} \pW_{\nu\rho\mu}, \nonumber \\
 \pW_{\mu\nu\rho\sigma} = (-1)^{1-\vec{a}_{\mu}\cdot\vec{a}_{\nu}} \pW_{\nu\mu\rho\sigma}
 = (-1)^{\vec{a}_{\mu}\cdot( \vec{a}_{\nu} + \vec{a}_{\rho})}
 \pW_{\nu\rho\mu\sigma}
 =-(-1)^{\vec{a}_{\mu}\cdot( \vec{a}_{\nu} + \vec{a}_{\rho} + \vec{a}_{\sigma})}
 \pW_{\nu\rho\sigma\mu}. \label{extension1}
\end{gather}
We further need to check the closure of multi-index matrices
\begin{align}
\llbracket \pX_{\mu \nu}, \pY_{\rho \sigma} \rrbracket &= \pX_{\mu \nu} \pY_{\rho \sigma} -(-1)^{(\vec{a}_{\mu}+\vec{a}_{\nu})\cdot (\vec{a}_{\rho}+\vec{a}_{\sigma})} \pY_{\rho \sigma} \pX_{\mu \nu} \nonumber \\
&=-{\rm i}^{-\vec{a}_{\mu}\cdot\vec{a}_{\nu}-\vec{a}_{\rho}\cdot\vec{a}_{\sigma}} \gamma_{\mu}\gamma_{\nu}\gamma_{\rho}\gamma_{\sigma} \otimes [X_{0},Y_{0}]
 = -{\rm i}^{-\vec{a}_{\mu}\cdot\vec{a}_{\nu}-\vec{a}_{\rho}\cdot\vec{a}_{\sigma}} \pZ_{\mu\nu\rho\sigma},
 \label{extension2}
\end{align}
where $\pZ_{\mu\nu\rho\sigma}$ is understood as in \eqref{extension1} and \eqref{extension2},
\begin{align}
 \llbracket \pX_{\mu \nu}, \pY_{\rho \sigma \tau} \rrbracket &= \pX_{\mu \nu} \pY_{\rho \sigma \tau} -(-1)^{(\vec{a}_{\mu}+\vec{a}_{\nu})\cdot (\vec{a}_{\rho}+\vec{a}_{\sigma}+\vec{a}_{\tau})} \pY_{\rho \sigma \tau} \pX_{\mu \nu} \nonumber \\
 &= -{\rm i}^{-\vec{a}_{\mu}\cdot\vec{a}_{\nu}}\gamma_{\mu}\gamma_{\nu}\gamma_{\rho}\gamma_{\sigma}\gamma_{\tau} \otimes [X_{0},Y_{1}]
 =-{\rm i}^{1-\vec{a}_{\mu}\cdot\vec{a}_{\nu}}\gamma_{\mu}\gamma_{\nu}\gamma_{\rho}\gamma_{\sigma}\gamma_{\tau} \otimes Z_0. \label{ClosureCl6no2}
\end{align}
There are five gamma matrices in this case so that one or two pairs of identical gamma matrices exist. When there exist one pair of identical matrices, say $ \gamma_{\nu} = \gamma_{\tau}$, \eqref{ClosureCl6no2} equals to $\pZ_{\mu\rho\sigma}$ up to a~constant factor. When there exist two pair of identical matrices, say $ \gamma_{\mu} = \gamma_{\rho} $ and $ \gamma_{\nu} = \gamma_{\sigma}$, \eqref{ClosureCl6no2}~equals to $\pZ_{\tau}$ up to a constant multiple. In this way, we see the closure of \eqref{ClosureCl6no2}.
Similarly, there exist identical gamma matrices in the following (anti)commutator
\begin{align}
\llbracket \pX_{\mu \nu \rho}, \pY_{\sigma \tau \lambda} \rrbracket &= \pX_{\mu \nu \rho} \pY_{\sigma \tau \lambda} -(-1)^{(\vec{a}_{\mu}+\vec{a}_{\nu}+\vec{a}_{\rho})\cdot (\vec{a}_{\sigma}+\vec{a}_{\tau}+\vec{a}_{\lambda})} \pY_{\sigma \tau \lambda} \pX_{\mu \nu \rho} \nonumber \\
&= -\gamma_{\mu}\gamma_{\nu}\gamma_{\rho}\gamma_{\sigma}\gamma_{\tau}\gamma_{\lambda} \otimes \{ X_{1},Y_{1} \}
= - \gamma_{\mu}\gamma_{\nu}\gamma_{\rho}\gamma_{\sigma}\gamma_{\tau}\gamma_{\lambda} \otimes W_0. \label{ClosureCl6no3}
\end{align}
A special subcase of this is three pairs of identical gamma matrices
\begin{equation*}
 \{ \pX_{\mu\nu\rho}, \pX_{\mu\nu\rho} \} = \mathbb{I}_8 \otimes W_0 = \pW_{(0,0,0)}.
\end{equation*}
Except the special case, there always exist two pairs of identical gamma matrices.
Therefore, \eqref{ClosureCl6no3} equals to $ \pW_{\mu\nu}$ up to a constant factor.
We thus have proved the closure of (anti)com\-mu\-ta\-tors.

We observed that the $\Z{3}$-graded (anti)commutator is reduced to the one of superalgebra.
It follows from this fact that a $\Z{3}$-graded Jacobi relation is also reduced to the one of superalgebra. Therefore, it is straightforward to verify that \eqref{Cl6real2} satisfies the $\Z{3}$-graded Jacobi relations.

Now we are able to use \eqref{Cl6real2} to construct a ${\rm Cl}(6)$ model of $\Z{3}$-graded SCQM. Taking $\mathfrak{s}$ as any model of SCQM, \eqref{Cl6real2} produces a corresponding model of $\Z{3}$-graded SCQM.
As a simplest example, here we take $ \mathfrak{s} = \mathfrak{osp}(1|2)$ given in \eqref{ospSCM}.
Then \eqref{Cl6real2} gives us 40 operators which close in a $\Z{3}$-graded extension of $ \mathfrak{osp}(1|2)$.
We denote this algebra simply by ${\mathcal G}_2$:
\begin{gather}
\pmb{X}_{(0,0,0)} = \mathbb{I}_8 \otimes X,\nonumber \\
\pmb{Q}_{\mu} = \gamma_{\mu} \otimes Q ,
 \qquad \pmb{S}_{\mu} = \gamma_{\mu} \otimes S ,\nonumber \\
\pmb{X}_{\mu \nu} = {\rm i}^{1-\vec{a}_{\mu}\cdot\vec{a}_{\nu}}\gamma_{\mu}\gamma_{\nu} \otimes X,\qquad \mu<\nu,
\nonumber \\
\pmb{Q}_{\mu\nu\rho} = {\rm i}\gamma_{\mu}\gamma_{\nu}\gamma_{\rho} \otimes Q,
\qquad \pmb{S}_{\mu\nu\rho} = {\rm i} \gamma_{\mu}\gamma_{\nu}\gamma_{\rho} \otimes S,
\qquad \mu < \nu < \rho, \nonumber \\
\pmb{X}_{0123} = {\rm i}\gamma_{1}\gamma_{2}\gamma_{3} \otimes X,\qquad X=H,D,K.
\label{Cl6modelNo2}
\end{gather}
The number of operator is double of the ${\rm Cl}(4)$ model discussed in Section~\ref{SubSEC:CL4} where we have 20 operators.
The reason of this difference is same as the $\Z{n}$-graded SQM considered in \cite{AAd2} and it is best seen in the next example:
\begin{gather*}
 [ \pmb{Q}_1, \pmb{Q}_2 ] = 2\gamma_1 \gamma_2 \otimes H = -2{\rm i} \pmb{H}_{12},
 \qquad
 \{\pmb{Q}_0, \pmb{Q}_3 \} = 2 \gamma_3 \otimes H = 2 \pmb{H}_{03}.
\end{gather*}
These are the relations of ${\rm Cl}(6)$ model and $ \deg(\pmb{H}_{12}) = \deg (\pmb{H}_{03}) = (1,1,0)$. As $ \gamma_1 \gamma_2 \neq \gamma_3$, $ \pmb{H}_{12} $~and $~\pmb{H}_{03}$ are
linearly independent.
The corresponding relations in the ${\rm Cl}(4)$ model are
\begin{gather*}
[\pmb{Q}_{100}, \pmb{Q}_{010}] = 2\gamma_1 \gamma_2 \otimes H,
\qquad
 \{\pmb{Q}_{111}, \pmb{Q}_{001} \} = 2{\rm i} \gamma_1 \gamma_2 \otimes H.
\end{gather*}
Obviously, the operators on the right hand side are not linearly independent.
Namely, in the~${\rm Cl}(4)$ model degeneracy of operators, which are linearly independent operators in ${\rm Cl}(6)$ model, happens and the number of the operators are reduced.

By using the explicit form of \eqref{Cl6modelNo2}, it is not difficult to see that the space $ \mathscr{H} = F(\mathbb{R})\otimes \mathbb{C}^{16} $ is not decoupled into two subspaces by the action of ${\mathcal G}_2$. This is a sharp contrast to the model in Section~\ref{SEC:Cl2n} and suggests the $\Z{3}$-graded SCQM \eqref{Cl6modelNo2} gives an irreducible representation of ${\mathcal G}_2$.
More precise analysis of irreducible representation of ${\mathcal G}_2$ will be done in a way similar to \cite{KANA} but it is beyond the scope of the present work.

Let us briefly analyse the spectrum of the ${\rm Cl}(6)$ model by employing the standard prescription of conformal mechanics.
That is, we define the operators
\begin{alignat*}{3}
 &\pmb{R}_{(0,0,0)}= \pmb{H}_{(0,0,0)} + \pmb{K}_{(0,0,0)},
 \qquad&&
 \pmb{L}_{(0,0,0)}^{\pm}= \frac{1}{2}( \pmb{K}_{(0,0,0)}- \pmb{H}_{(0,0,0)}) \pm {\rm i} \pmb{D}_{(0,0,0)},&\\
& \pmb{a}_{\mu}= \pmb{S}_{\mu} + {\rm i} \pmb{Q}_{\mu},
 \qquad &&
 \pmb{a}_{\mu}^{\dagger}= \pmb{S}_{\mu} -{\rm i} \pmb{Q}_{\mu},&
 \\
 &\pmb{a}_{\mu\nu\rho}= \pmb{S}_{\mu\nu\rho} + {\rm i} \pmb{Q}_{\mu\nu\rho},
 \qquad &&
 \pmb{a}_{\mu\nu\rho}^{\dagger}= \pmb{S}_{\mu\nu\rho} -{\rm i} \pmb{Q}_{\mu\nu\rho},&
\end{alignat*}
and take the eigenspace of $ \pmb{R}_{(0,0,0)}$, which is $ L^2(\mathbb{R}) \otimes \mathbb{C}^{16}$, as the Hilbert space of the theory.
There exist twice many creation and annihilation operators than ${\rm Cl}(4)$ model
\begin{alignat*}{3}
 &\big[\pmb{R}_{(0,0,0)}, \pmb{a}_{\mu} \big]= - \pmb{a}_{\mu},
 \qquad&&
 \big[\pmb{R}_{(0,0,0)}, \pmb{a}^{\dagger}_{\mu} \big]= - \pmb{a}^{\dagger}_{\mu},&\\
& \big[\pmb{R}_{(0,0,0)}, \pmb{a}_{\mu\nu\rho} \big]= - \pmb{a}_{\mu\nu\rho},
 \qquad&&
 \big[\pmb{R}_{(0,0,0)}, \pmb{a}^{\dagger}_{\mu\nu\rho} \big]= - \pmb{a}^{\dagger}_{\mu\nu\rho},&
 \\
& \big[\pmb{R}_{(0,0,0)}, \pmb{L}_{(0,0,0)}^{\pm} \big]= \pm 2 \pmb{L}_{(0,0,0)}^{\pm}.\qquad &&&
\end{alignat*}
These creation and annihilation operators satisfy the relations similar to the ${\rm Cl}(4)$ model
\begin{gather*}
 \big\{ \pmb{a}_{\mu}, \pmb{a}_{\mu}^{\dagger} \big\}
 =
 \big\{ \pmb{a}_{\mu\nu\rho}, \pmb{a}_{\mu\nu\rho}^{\dagger} \big\} = 2 \pmb{R}_{(0,0,0)},
 \nonumber \\
 \big\{ \pmb{a}_{\mu}^{\dagger}, \pmb{a}_{\mu}^{\dagger} \big\}
 = \big\{ \pmb{a}_{\mu\nu\rho}^{\dagger}, \pmb{a}_{\mu\nu\rho}^{\dagger} \big\} = 2 \pmb{L}^+_{(0,0,0)},
 \qquad
 \{ \pmb{a}_{\mu}, \pmb{a}_{\mu} \}
 = \{ \pmb{a}_{\mu\nu\rho}, \pmb{a}_{\mu\nu\rho} \} = 2 \pmb{L}^-_{(0,0,0)}.
\end{gather*}
Furthermore, $ \pmb{a}_{\mu}$ and $ \pmb{a}_{\mu}^{\dagger} $ satisfy a Klein deformed oscillator algebra
\begin{gather*}
 \big[ \pmb{a}_{\mu}, \pmb{a}_{\mu}^{\dagger} \big]
 =
 \big[ \pmb{a}_{\mu\nu\rho}, \pmb{a}_{\mu\nu\rho}^{\dagger} \big] =
 \mathbb{I}_{16} - 2\beta F,
 \qquad F:= \mathbb{I}_8 \otimes \sigma_3,
 \nonumber \\
 \{ F, \pmb{a}_{\mu} \} = \big\{ F, \pmb{a}_{\mu}^{\dagger} \big\}
 = \{ F, \pmb{a}_{\mu\nu\rho} \} = \big\{ F, \pmb{a}_{\mu\nu\rho}^{\dagger} \big\}
 = 0,
 \qquad F^2 = \mathbb{I}_{16}.
\end{gather*}
With these relations one may have
\begin{gather*}
 \pmb{R}_{(0,0,0)} = \pmb{a}_{\mu}^{\dagger} \pmb{a}_{\mu} + \frac{1}{2} (\mathbb{I}_{16}-2\beta F)
 =
 \pmb{a}_{\mu\nu\rho}^{\dagger} \pmb{a}_{\mu\nu\rho} + \frac{1}{2} (\mathbb{I}_{16}-2\beta F).
\end{gather*}
Thus the ground state $ \Psi(x)$ or $\pmb{R}_{(0,0,0)}$ is defined by
\begin{equation}
 \pmb{a}_{\mu} \Psi(x)= \pmb{a}_{\mu \nu \rho} \Psi(x) = 0. \label{Cl6DefGroundState}
\end{equation}
We write $\Psi(x)$ in components
\begin{gather*}
 \Psi(x) =
(\psi_{000},\psi_{111},\psi_{110},\psi_{001},\psi_{011},\psi_{100},\psi_{101},\psi_{010},\psi_{110},\psi_{001},\psi_{000},\\
\hphantom{\Psi(x) =(}{} \psi_{111},\psi_{101},\psi_{010},\psi_{011},\psi_{100})^{\rm T}.
\end{gather*}
Then the condition \eqref{Cl6DefGroundState} gives the same relations as ${\rm Cl}(4)$ model \eqref{GScondition1} and \eqref{GScondition2}.
It follows that the ground state of ${\rm Cl}(6)$ model is eight-fold degenerate.
The excited states are obtained by repeated application of $ \pmb{a}_{\mu}^{\dagger} $ and $ \pmb{a}_{\mu\nu\rho}^{\dagger} $ on the ground state.
Repeating the argument same as the ${\rm Cl}(4)$ model, one may see that the spectrum of $\pmb{R}_{(0,0,0)}$ is equally spacing and the excited state has eight-fold degeneracy.

As an abstract Lie algebra we regard ${\mathcal G}_2$ is inequivalent to ${\mathcal G}_1$ as
$ \dim {\mathcal G}_2 = 40 > \dim {\mathcal G}_1$.

\section{Concluding remarks} \label{SEC:CR}

\looseness=-1 We showed that many models of SCQM are able to extend to $\Z{n}$-graded setting
by the use of the Clifford algebras ${\rm Cl}(2(n-1))$ and $ {\rm Cl}(2n)$.
It was also shown the existence of a sequence of models of $\Z{3}$-graded $\mathfrak{osp}(1|2)$ SCQM and we analyzed the spectrum of the models.
Most likely, for a~given model of standard SCQM there would exists a sequence of models of $\Z{n}$-graded SCQM produced via the sequence of the Clifford algebras~\eqref{ZnSeq}.
However, full analysis of $\Z{n}$-graded SCQM will require lengthy computation and invention of better notations (especially for the models via higher-dimensional Clifford algebras)
which make the presentation simpler and more readable.
Therefore, we are planning to present them in a separate publication.
We convince that the present analysis of $\Z{3}$-graded SCQM provides all the essentials of $\Z{n}$-graded extensions.

Although the existence of $\Z{n}$-graded SCQM has been established,
its physical implications and how much it differs from the standard SCQM are not clear yet.
To have better understanding of~$\Z{3}$ and higher graded SCQM, there would be some more works to be done.
For example, one may consider multiparticle extensions of the models presented in this paper. As showed in $\Z{2}$-graded SQM, difference from the standard SQM becomes clear when a multiparticle model is considered.
A multiparticle extension may be done in a way similar to~\cite{Topp}.

The second example is classical theories of $\Z{3}$-graded SCQM which reproduce the models of this work upon quantization.
Such classical theories will shed a new light on $\Z{3}$-graded SCQM and they have their own interest.
For the simpler grading by~$ \Z{2}$,
$D$-module presentation and superfield approach to the classical theory of $\Z{2}$-graded SQM are discussed in the literature~\mbox{\cite{AKTcl,brusigma}}.
It is a very interesting but challenging problem to generalize these to $\Z{n}$-graded $ (n \geq 3)$ setting since integration on $\Z{n}$-graded $ (n \geq 3)$ supermanifolds has not been established yet~\cite{Pon}.
Nonlinear realization is a widely used approach to superconformal mechanics, see, e.g., \cite{BelKri, FedIvaLec}. $ \Z{3}$-graded extension of nonlinear realization will be possible and it will give some geometrical understanding of $\Z{3}$-graded SCQM.

\appendix

\section[Definition of Z2n-graded Lie superalgebra]{Definition of $\boldsymbol{\Z{n}}$-graded Lie superalgebra}

In this appendix we give a rigorous definition of $\Z{n}$-graded Lie superalgebra \cite{Ree,rw1,rw2}.
Let~$\g$ be a vector space over $\mathbb{R}$ or $ \mathbb{C}$ and $\vec{a}=(a_1,a_2,\dots,a_n)$ an element of $\Z{n}$.
Suppose that $\g$ is a~direct sum of graded subspaces labelled by $\vec{a}$
\begin{equation*}
 \g = \bigoplus_{\vec{a}} \g_{\vec{a}}.
\end{equation*}
Homogeneous elements of $\g_{\vec{a}}$ are denoted by $ X_{\vec{a}}, Y_{\vec{a}}, \dots $.
If $\g$ admits a bilinear operation (the general Lie bracket), denoted by
$ \llbracket \cdot, \cdot \rrbracket$, satisfying the identities
\begin{gather}
 \llbracket X_{\vec{a}}, Y_{\vec{b}} \rrbracket \in \g_{\vec{a}+\vec{b}},\qquad
 \llbracket X_{\vec{a}}, Y_{\vec{b}} \rrbracket = -(-1)^{\vec{a}\cdot\vec{b}} \llbracket Y_{\vec{b}}, X_{\vec{a}} \rrbracket,\nonumber \\
 (-1)^{\vec{a}\cdot\vec{c}} \llbracket X_{\vec{a}}, \llbracket Y_{\vec{b}}, Z_{\vec{c}} \rrbracket \rrbracket
 +
 (-1)^{\vec{b}\cdot\vec{a}} \llbracket Y_{\vec{b}}, \llbracket Z_{\vec{c}}, X_{\vec{a}} \rrbracket \rrbracket
 +
 (-1)^{\vec{c}\cdot\vec{b}} \llbracket Z_{\vec{c}}, \llbracket X_{\vec{a}}, Y_{\vec{b}} \rrbracket \rrbracket
 = 0, \label{GJdef}
\end{gather}
where
\begin{equation*}
 \vec{a} + \vec{b} = (a_1+b_1, a_2+b_2, \dots) \in \Z{n},
 \qquad
 \vec{a}\cdot \vec{b} = \sum_{k=1}^n a_k b_k,
\end{equation*}
then $\g$ is referred to as a $\Z{n}$-graded Lie superalgebra.
The relation \eqref{GJdef} is called the $\Z{n}$-graded Jacobi relation.

We take $\g$ to be contained in its enveloping algebra, via the identification
\begin{equation}
 \llbracket X_{\vec{a}}, Y_{\vec{b}} \rrbracket = X_{\vec{a}} Y_{\vec{b}} - (-1)^{\vec{a}\cdot \vec{b}}
Y_{\vec{b}} X_{\vec{a}}, \label{GLB}
\end{equation}
where an expression such as $ X_{\vec{a}} Y_{\vec{b}}$ is understood to denote the associative product
on the enveloping algebra.
In other words, by definition, in the enveloping algebra the general Lie bracket $ \llbracket \cdot, \cdot
\rrbracket $ for homogeneous elements coincides with either a commutator or anticommutator.

 This is a natural generalization of Lie superalgebra which is defined on $\mathbb{Z}_2$-grading structure. Namely, the vector $\vec{a}$ is one-dimensional
\begin{equation*}
 \g = \g_{(0)} \oplus \g_{(1)}
\end{equation*}
with
$ \vec{a} + \vec{b} = (a+b)$, $\vec{a} \cdot \vec{b} = ab$.

\subsection*{Acknowledgements}

The authors would like to thank the referees for the valuable comments for improvement of this paper.

\pdfbookmark[1]{References}{ref}
\LastPageEnding


\begin{thebibliography}{99}
\footnotesize\itemsep=0pt

\bibitem{NAcl}
Aizawa N., Generalization of superalgebras to color superalgebras and their
 representations, \href{https://doi.org/10.1007/s00006-018-0847-x}{\textit{Adv. Appl. Clifford Algebr.}} \textbf{28} (2018), 28,
 14~pages, \href{https://arxiv.org/abs/1712.03008}{arXiv:1712.03008}.

\bibitem{AAd2}
Aizawa N., Amakawa K., Doi S., {$\mathbb{Z}_2^n$}-graded extensions of
 supersymmetric quantum mechanics via {C}lifford algebras, \href{https://doi.org/10.1063/1.5144325}{\textit{J.~Math.
 Phys.}} \textbf{61} (2020), 052105, 13~pages, \href{https://arxiv.org/abs/1912.11195}{arXiv:1912.11195}.

\bibitem{AAD}
Aizawa N., Amakawa K., Doi S., {$\mathcal N$}-extension of double-graded
 supersymmetric and superconformal quantum mechanics, \href{https://doi.org/10.1088/1751-8121/ab661c}{\textit{J.~Phys.~A:
 Math. Theor.}} \textbf{53} (2020), 065205, 14~pages, \href{https://arxiv.org/abs/1905.06548}{arXiv:1905.06548}.

\bibitem{ACKT}
Aizawa N., Cunha I.E., Kuznetsova Z., Toppan F., On the spectrum-generating
 superalgebras of the deformed one-dimensional quantum oscillators,
 \href{https://doi.org/10.1063/1.5085164}{\textit{J.~Math. Phys.}} \textbf{60} (2019), 042102, 18~pages,
 \href{https://arxiv.org/abs/1812.00873}{arXiv:1812.00873}.

\bibitem{AKTT1}
Aizawa N., Kuznetsova Z., Tanaka H., Toppan F.,
 {$\mathbb{Z}_2\times\mathbb{Z}_2$}-graded {L}ie symmetries of the
 {L}\'evy-{L}eblond equations, \href{https://doi.org/10.1093/ptep/ptw176}{\textit{Prog. Theor. Exp. Phys.}} \textbf{2016} (2016),
 123A01, 26~pages, \href{https://arxiv.org/abs/1609.08224}{arXiv:1609.08224}.

\bibitem{AKTT2}
Aizawa N., Kuznetsova Z., Tanaka H., Toppan F., Generalized supersymmetry and {L}\'evy-{L}eblond equation, in Physical and Mathematical Aspects of Symmetries,
Editors J.-P.~Gazeau, S.~Faci, T.~Micklitz, R.~Scherer, F.~Toppan, \href{https://doi.org/10.1007/978-3-319-69164-0_11}{Springer}, Cham, 2017, 79--84, \href{https://arxiv.org/abs/1609.08760}{arXiv:1609.08760}.


\bibitem{AKT}
Aizawa N., Kuznetsova Z., Toppan F., The quasi-nonassociative exceptional
 {$F(4)$} deformed quantum oscillator, \href{https://doi.org/10.1063/1.5016915}{\textit{J.~Math. Phys.}} \textbf{59}
 (2018), 022101, 13~pages, \href{https://arxiv.org/abs/1711.02923}{arXiv:1711.02923}.

\bibitem{AKTcl}
Aizawa N., Kuznetsova Z., Toppan F., ${\mathbb Z}_2\times {\mathbb Z}_2$-graded
 mechanics: the classical theory, \href{https://doi.org/10.1140/epjc/s10052-020-8242-x}{\textit{Eur. Phys.~J.~C}} \textbf{80} (2020),
 668, 14~pages, \href{https://arxiv.org/abs/2003.06470}{arXiv:2003.06470}.

\bibitem{AKTqu}
Aizawa N., Kuznetsova Z., Toppan F., {$\mathbb Z_2\times \mathbb Z _2$}-graded
 mechanics: the quantization, \href{https://doi.org/10.1016/j.nuclphysb.2021.115426}{\textit{Nuclear Phys.~B}} \textbf{967} (2021),
 115426, 30~pages, \href{https://arxiv.org/abs/2021.11542}{arXiv:2021.11542}.

\bibitem{KANA}
Amakawa K., Aizawa N., A classification of lowest weight irreducible modules
 over {$\mathbb{Z}_2^2$}-graded extension of {$\mathfrak{osp}(1|2)$}, \href{https://doi.org/10.1063/5.0037493}{\textit{J.~Math.
 Phys.}} \textbf{62} (2021), 043502, 18~pages, \href{https://arxiv.org/abs/2011.03714}{arXiv:2011.03714}.

\bibitem{BelKri}
Bellucci S., Krivonos S., Supersymmetric mechanics in superspace, in
 Supersymmetric Mechanics, {V}ol.~1, \textit{Lecture Notes in Phys.}, Vol.~698, \href{https://doi.org/10.1007/3-540-33314-2_2}{Springer}, Berlin, 2006, 49--96, \href{https://arxiv.org/abs/hep-th/0602199}{arXiv:hep-th/0602199}.

\bibitem{Bruce}
Bruce A.J., On a $\mathbb{Z}_2^n$-graded version of supersymmetry,
 \href{https://doi.org/10.3390/sym11010116}{\textit{Symmetry}} \textbf{11} (2019), 116, 20~pages, \href{https://arxiv.org/abs/1812.02943}{arXiv:1812.02943}.

\bibitem{brusigma}
Bruce A.J., {$\mathbb Z_2 \times \mathbb Z_2$}-graded supersymmetry: 2-d sigma
 models, \href{https://doi.org/10.1088/1751-8121/abb47f}{\textit{J.~Phys.~A: Math. Theor.}} \textbf{53} (2020), 455201,
 25~pages, \href{https://arxiv.org/abs/2006.08169}{arXiv:2006.08169}.

\bibitem{BruDup}
Bruce A.J., Duplij S., Double-graded supersymmetric quantum mechanics,
 \href{https://doi.org/10.1063/1.5118302}{\textit{J.~Math. Phys.}} \textbf{61} (2020), 063503, 13~pages,
 \href{https://arxiv.org/abs/1904.06975}{arXiv:1904.06975}.

\bibitem{CaRoTo}
Carrion H.L., Rojas M., Toppan F., Quaternionic and octonionic spinors. {A}
 classification, \href{https://doi.org/10.1088/1126-6708/2003/04/040}{\textit{J.~High Energy Phys.}} \textbf{2003} (2003), 040,
 28~pages, \href{https://arxiv.org/abs/hep-th/0302113}{arXiv:hep-th/0302113}.

\bibitem{DFF}
de~Alfaro V., Fubini S., Furlan G., Conformal invariance in quantum mechanics,
 \href{https://doi.org/10.1007/BF02785666}{\textit{Nuovo Cimento~A}} \textbf{34} (1976), 569--612.

\bibitem{FedIvaLec}
Fedoruk S., Ivanov E., Lechtenfeld O., Superconformal mechanics,
 \href{https://doi.org/10.1088/1751-8113/45/17/173001}{\textit{J.~Phys.~A: Math. Theor.}} \textbf{45} (2012), 173001, 59~pages,
 \href{https://arxiv.org/abs/1112.1947}{arXiv:1112.1947}.

\bibitem{LR}
Le~Roy B., $\mathbb{Z}_n^3$-graded colored supersymmetry,
 \href{https://doi.org/10.1023/A:1021491927893}{\textit{Czechoslovak~J. Phys.}} \textbf{47} (1997), 47--54,
 \href{https://arxiv.org/abs/hep-th/9608074}{arXiv:hep-th/9608074}.

\bibitem{Oka}
Okazaki T., Superconformal quantum mechanics from M2-branes, Ph.D.~Thesis,
 {C}alifornia Institute of Technology, USA, Osaka University, Japan,
 \href{https://arxiv.org/abs/1503.03906}{arXiv:1503.03906}.

\bibitem{Okubo}
Okubo S., Real representations of finite {C}lifford algebras.
 {I}.~{C}lassification, \href{https://doi.org/10.1063/1.529277}{\textit{J.~Math. Phys.}} \textbf{32} (1991),
 1657--1668.

\bibitem{Pa}
Papadopoulos G., Conformal and superconformal mechanics, \href{https://doi.org/10.1088/0264-9381/17/18/310}{\textit{Classical
 Quantum Gravity}} \textbf{17} (2000), 3715--3741, \href{https://arxiv.org/abs/hep-th/0002007}{arXiv:hep-th/0002007}.

\bibitem{Pon}
Poncin N., Towards integration on colored supermanifolds, in Geometry of Jets
 and Fields, \textit{Banach Center Publ.}, Vol.~110, \href{https://doi.org/10.4064/bc110-0-14}{Polish Acad. Sci. Inst.
 Math.}, Warsaw, 2016, 201--217.

\bibitem{Ree}
Ree R., Generalized {L}ie elements, \href{https://doi.org/10.4153/CJM-1960-044-x}{\textit{Canadian~J. Math.}} \textbf{12}
 (1960), 493--502.

\bibitem{rw1}
Rittenberg V., Wyler D., Generalized superalgebras, \href{https://doi.org/10.1016/0550-3213(78)90186-4}{\textit{Nuclear Phys.~B}}
 \textbf{139} (1978), 189--202.

\bibitem{rw2}
Rittenberg V., Wyler D., Sequences of {$Z_{2}\oplus Z_{2}$} graded {L}ie
 algebras and superalgebras, \href{https://doi.org/10.1063/1.523552}{\textit{J.~Math. Phys.}} \textbf{19} (1978),
 2193--2200.

\bibitem{sch1}
Scheunert M., Generalized {L}ie algebras, \href{https://doi.org/10.1063/1.524113}{\textit{J.~Math. Phys.}} \textbf{20}
 (1979), 712--720.

\bibitem{tol2}
Tolstoy V.N., Super-de {S}itter and alternative super-{P}oincar\'e symmetries,
 in Lie Theory and its Applications in Physics, \textit{Springer Proc. Math.
 Stat.}, Vol.~111, \href{https://doi.org/10.1007/978-4-431-55285-7_26}{Springer}, Tokyo, 2014, 357--367, \href{https://arxiv.org/abs/1610.01566}{arXiv:1610.01566}.

\bibitem{Topp}
Toppan F., {$\mathbb Z_2 \times \mathbb Z_2$}-graded parastatistics in
 multiparticle quantum {H}amiltonians, \href{https://doi.org/10.1088/1751-8121/abe2f2}{\textit{J.~Phys.~A: Math. Theor.}}
 \textbf{54} (2021), 115203, 35~pages, \href{https://arxiv.org/abs/2008.11554}{arXiv:2008.11554}.

\end{thebibliography}
\end{document}